\documentclass[twocolumn,preprintnumbers,amsmath,amssymb]{revtex4}
\usepackage{graphicx}
\usepackage{bm}% bold math

\begin{document}
\title {Parity effect in Al and Nb single electron transistors in a tunable environment}

\author{A.\ M.\ Savin}
\author{M.\ Meschke}
\author{J.\ P.\ Pekola}
\affiliation{Low Temperature Laboratory, Helsinki University of
Technology, P.O. Box 3500, FIN-02015 TKK, Finland}
\author{Yu.\ A.\ Pashkin}
\altaffiliation[on leave from ]{Lebedev Physical Institute, Moscow
119991, Russia}

\author{T.\ F.\ Li}
\altaffiliation {Institute of Microelectronics, Tsinghua University,
Beijing 100084, China}
\author{H.\ Im}
\altaffiliation{Department of Semiconductor Science, Dongguk
University, Phil-Dong, Seoul 100-715, Korea}

\author{J.\ S.\ Tsai}
\affiliation{Nano Electronics Research Laboratories, NEC
Corporation, 34 Miyukigaoka, Tsukuba, Ibaraki 305-8501, Japan and
The Institute of Physical and Chemical Research, 2-1 Hirosawa, Wako,
Saitama 351-0198, Japan}

\begin{abstract}
Two different types of Cooper pair transistors, with Al and Nb
islands, have been investigated in a tunable electromagnetic
environment. The device with an Al island demonstrates gate charge
modulation with $2e$-periodicity in a wide range of environmental
impedances at bath temperatures below 340 mK. Contrary to the
results of the Al sample, we were not able to detect
$2e$-periodicity under any conditions on similar samples with Nb
island. We attribute this to the material properties of Nb.
\end{abstract}

\maketitle

A Cooper pair transistor (CPT) is a basic element for a number of
applications such as ultrasensitive electrometry, quantum computing
and metrology. Until now Al, with superior characteristics with Al
oxide as the tunnel barrier, has been the material of choice for a
CPT. Yet, due to its higher superconducting gap, $\Delta$, as
compared to that of Al, Nb would be an interesting alternative for
superconducting devices. Larger $\Delta$ ensures a wider range of
operation in terms of the working temperature and tolerance to
external noise. Moreover, the operation speed is typically
proportional to the value of $\Delta$.

A major, still largely unexplained, disadvantage in employing CPT
based devices is their susceptibility to quasiparticle poisoning.
Ideally, only paired electrons contribute to the charge transport in
CPTs and the island parity remains preserved resulting in
$2e$-periodicity of the CPT transport. However, in a real experiment
single-electron, or quasiparticle, tunneling may change the parity
and transport periodicity. This is the effect usually referred to as
quasiparticle poisoning. In Al based devices quasiparticle poisoning
can be suppressed in many cases \cite{Saclay, Aumentado, Yamamoto,
Ferguson, Corlevi}, but there are no reports on quasiparticle-free
CPTs made of Nb, although a wealth of experiments already exist on
these systems \cite{Harada, Patel, Dolata, Kim, WatanabeNb,Im}. This
is surprising to some extent because larger $\Delta$ should in
principle diminish quasiparticle poisoning. Moreover, in the case of
CPT with Al leads and a Nb island the larger superconducting gap of
the island should further suppress quasiparticle tunneling into the
island \cite{Aumentado, Yamamoto, Ferguson}. The question remains
whether the quasiparticle poisoning in Nb structures is due to their
susceptibility to environment fluctuations, or whether it is an
intrinsic material property of Nb. The aim of the present work is to
investigate the parity effect under identical experimental
conditions in CPTs with Al leads but with either Al or Nb island. We
also employ the recently developed concept of tunable environment
\cite{Watanabe, Corlevi}, which should significantly suppress
quasiparticle poisoning \cite{Corlevi}.

\begin{figure}[b!]
\begin{center}
\includegraphics[width=8.5cm,clip]{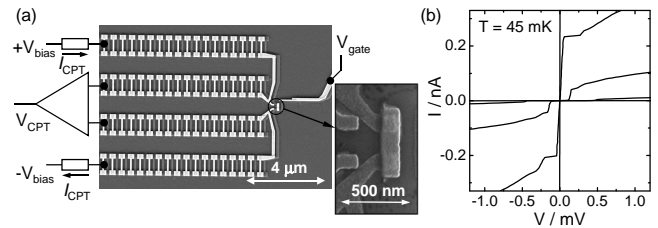}
\end{center}
\caption{(a) Electron micrograph of a CPT with SQUID arrays and
diagram of measurement circuit (left) and an enlarged image of the
CPT island (right). (b) Current-voltage characteristics of two SQUID
arrays around zero bias in sample Al-6 at $T = 45$ mK for values of
magnetic flux from zero to half flux quantum through each SQUID.
Zero bias resistance of the arrays varies from 7 k$\Omega$ up to 5
G$\Omega$ in this case.}
%and the upper value is most likely limited by the finite impedance
%of the measuring circuit, not the arrays themselves.}
\label{Fig1_2a}
\end{figure}

The measured CPTs consist of a superconducting island (Al or Nb)
coupled to two superconducting Al leads via nominally identical
Josephson junctions and capacitively coupled to a gate electrode
(Fig. \ref{Fig1_2a}). The samples are fabricated by two angle
evaporation through a Ge suspended mask supported by a thermally
stable polymer \cite{Dubos}. In both cases the first evaporated
layer is Al and the tunnel barrier is formed by thermal oxidation of
Al. The CPT island (15 nm thick Al or 30 nm thick Nb) with
dimensions $460\times130$ nm$^{2}$ is connected to two Al leads (25
nm or 15 nm thick) by two tunnel junctions whose dimensions were
slightly varied around $100\times100$ nm$^{2}$ giving the charging
energy of the devices, $E_c = e^2/(2C_\Sigma)$, of about 100
$\mu$eV, $C_\Sigma$ being the total capacitance of the island. Close
to the CPT island, each of the two Al leads is split into two
22-$\mu$m-long SQUID arrays consisting of 81 SQUIDs each.

The measurements are performed in a $^{3}$He/$^{4}$He dilution
refrigerator in a four-probe configuration for CPTs and using only
two probes for SQUID arrays. All measurement lines were filtered
using 1.5 m of thermocoax between 1 K plate and the sample stage and
a low pass filter on the sample stage. SQUID arrays serve as
additional filters and allow to modify environmental impedance of
the CPT by variation of a perpendicular external magnetic field. The
parameters of the investigated samples are listed in Table
\ref{Samples}. The charging energy of the transistor was derived
based on Coulomb blockade at a temperature above the critical
temperature of both superconductors. The measured values of the
charging energy are in agreement with the charging energy derived
from the size of the junctions. Josephson coupling energy $E_J$ for
one CPT junction is derived from the normal state resistance of the
junction $R_{N}$ assuming $\Delta_{\text{Al}} \simeq 0.2$ meV and
$\Delta_{\text{Nb}} \simeq 1$ meV.

\begin{table}
 \centering
  \caption{Parameters of the measured CPTs.}\label{Samples}
\begin{tabular}{|p{2.1cm}|p{1.4cm}|p{1.4 cm}|p{1.4cm}|p{1.4cm}|}

\hline

  Sample & Nb-1&  Nb-4& Nb-7    & Al-6 \\

 \hline
  Island & Nb&  Nb& Nb    & Al \\

 \hline

 $2R_{N}$ (k$\Omega$)  &  24&   183&   112    &     63.2 \\

% \hline

% $I_{c} $ (nA) &  57 &7.4 & 3& 10\\

\hline

 $E_{J}$ ($\mu$eV) &  116&   15 &   25 &     21 \\

\hline

  $E_{C}$ ($\mu$eV)&  112  &   83  &  183    &     118\\

 \hline

  $R_{0}^{\rm min}$ (k$\Omega$) &  0.45 &   0.9 &   1.0    &     7\\

\hline

 $R_{0}^{\rm max}$ (G$\Omega$) &  0.75 &   2.1 &   11    &     5\\

 \hline

 %\hline

\end{tabular}
\end{table}

We characterize the environment simply by their zero bias resistance
$R_{0}$ bearing in mind that the real impedance may be different.
$R_{0}$ is obtained from the current-voltage characteristics (IVCs)
of the arrays. As an example, IVCs of two arrays in series for
sample Al-6 measured at different values of the magnetic field
threading the SQUID loops are shown in Fig. 1(b). The applied
magnetic field suppresses supercurrent of the SQUIDs, which leads to
increase of the zero bias resistance. At higher magnetic fields
Coulomb blockade becomes pronounced and develops in a wider voltage
range. Maximum and minimum values of $R_{0}$ of two SQUID arrays
connected in series are given in Table \ref{Samples}. The minimum
$R_{0}$ and its dynamic range vary from sample to sample, and this
can be ascribed to different values of critical currents from sample
to sample and to the spread in SQUID parameters. In the Al/Nb hybrid
samples the SQUID junctions are formed between two different
superconductors, which also affects their characteristics.
Nevertheless, it was possible to tune $R_{0}$ over almost six orders
of magnitude in all the samples.

\begin{figure}[t!]
\begin{center}
\includegraphics[width=8.5cm,clip]{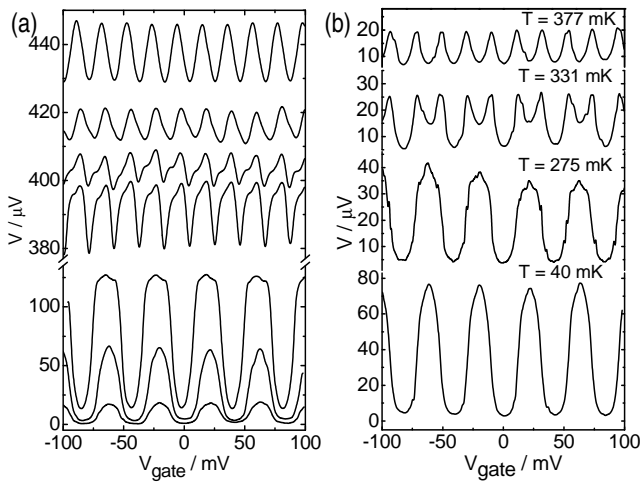}
\end{center}
\caption{Gate induced CPT voltage modulation of Al-6 sample at
$R_{0} = 7$ k$\Omega$ at different bias currents at $T=40$ mK (a)
and at different bath temperatures (b).} \label{FigGateAl}
\end{figure}
\indent The CPT is biased through the SQUID arrays which means that
the IVC and gate modulation are measured in the current biased
regime. General features of the measured all-aluminum sample (Al-6)
are in agreement with the theoretical predictions for a CPT
\cite{Corlevi}. At a high enough environmental impedance, Coulomb
blockade of Cooper pair tunneling develops. At $R_{0} > 10$
M$\Omega$ IVC demonstrates back bending (not shown), which is a
manifestation of Bloch oscillations. At $R_{0} < 10$ M$\Omega$ the
gate modulation of IVC is $2e$-periodic at the bias points
corresponding to the supercurrent branch and $e$-periodic at higher
current values (Fig. \ref{FigGateAl}a). At low voltage (supercurrent
branch) the modulation period is 42 mV, which is twice larger than
at higher voltages. The smaller period coincides with that observed
in the same CPT in the normal state (at high temperature or in high
magnetic field), which confirms that the observed reduction of
modulation period corresponds to the $2e-1e$ transition. In our case
$2e$-periodicity could be observed in the Al sample at all values of
the environmental impedance, unlike in the experiments reported
earlier \cite{Corlevi, Kuo}. Corlevi et al. \cite{Corlevi} observed
transitions from $1e$ to $2e$ only at rather high values of $R_0$
($>$ 5 M$\Omega$) using a similar Al CPT. This may reflect a
difference in the impedance seen by the CPT for the same value of
$R_0$ due to the different layout in our experiment and in ref.
\cite{Corlevi} or filtering of the signal lines. This is consistent
with the relatively high effective noise temperature (150 mK)
reported in \cite{Corlevi}. Also, energy profile in our Al CPT (15
nm thick island and 25 nm thick leads) may be favorable for the
observation of $2e$-periodicity \cite{Yamamoto}. Our results for
high impedance regime (IVC with negative slope) are similar to those
in Ref. \cite{Corlevi}: $2e$-modulation was observed in the Bloch
regime at low bias currents, and $e$-periodicity in the Zener
tunneling regime. Gate modulation of sample Al-6 at different bath
temperatures and at low array impedance of $R_{0} = 7$ k$\Omega$ is
presented in Fig. \ref{FigGateAl}b. Increase of temperature leads to
increased concentration of thermal quasiparticles and as a
consequence to a $2e\rightarrow e$ transition. Crossover temperature
$T^{*}$ for Al-6 sample is about 340 mK, which agrees with the
theoretical prediction $T^*=\Delta / [k_{B}\ln(N_{eff})]$
\cite{Tuominen}, where $N_{eff}$ is the number of quasiparticle
states on the island available for thermal excitation.
%Our $T^*$ is
%slightly higher than that for the Al CPT with comparable dimensions
%in \cite{Corlevi} that may be related to a difference in $\Delta$
%and the island volume.
\newline
\begin{figure}[t!]
\begin{center}
\includegraphics[width=8.5cm,clip]{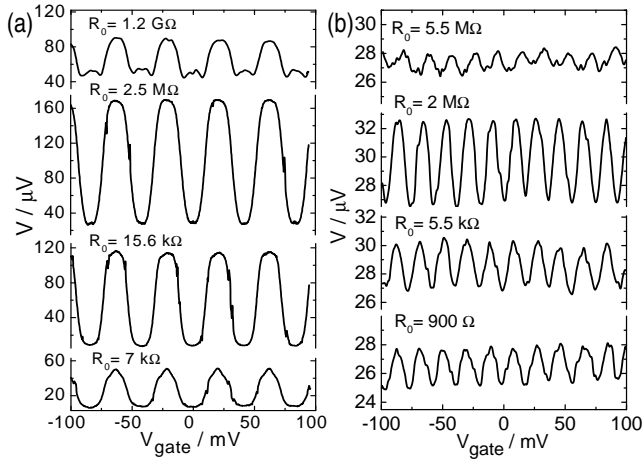}
\end{center}
\caption{Gate induced CPT voltage modulation of sample Al-6 (a) and
Nb-4 (b) at different magnetic fields and at $T = 40$ mK. Biasing
point for both samples corresponds to supercurrent branch of IVC.
Zero bias resistance of the SQUID arrays is marked on the curves.}
\label{GateModNb}
\end{figure}
\indent Three Nb samples with different $E_{J}/E_{C}$ ratios were
measured (see Table \ref{Samples}). The parameters and measured
characteristics of sample Nb-4 are rather similar to those of sample
Al-6 described above. Samples with larger (Nb-1) and lower (Nb-7)
$E_{J}/E_{C}$ ratio were also investigated. All samples were
measured over a wide range of environmental impedances and biasing
currents. Like in the Al sample, Coulomb blockade of the hybrid
samples becomes more pronounced for higher environmental impedance.
Gate modulation is, however, significantly weaker in all Nb samples
as compared to that in Al-6. Figure \ref{GateModNb} gives a
comparison of the gate modulation curves for Al-6 (a) and Nb-4 (b)
samples at $T = 40$ mK. They were recorded in the supercurrent
branch of the CPT and at different values of $R_0$ of the arrays.
Initially, as expected, the voltage amplitude in the gate modulation
increases with $R_0$, and then it drops in the back-bending regime
of the CPT. Under all experimental conditions the period of gate
induced modulation in Nb samples, including the normal state, is
about 20 mV, which is about the same as the $e$ period of sample
Al-6. We thus conclude that all our samples with Nb island exhibit
only $e$-periodic modulation. To explain this, one should address
material properties of Nb \cite{Halbritter}. We believe the observed
strong quasiparticle poisoning and a large subgap leakage in either
Al/Nb or Nb/Nb junctions, as compared to all-Al junctions, have the
same origin. With the angle evaporation technique, we obtain the
ratio of subgap to normal state resistance for all-All junctions of
the order of 1000 while it is only about 50 for all-Nb junctions
with a Nb-oxide barrier \cite{ImJVST}. In the case of the hybrid
structures studied in this work, the ratio is equal to 200, similar
to that measured in tri-layer Josephson junctions. The reason for
the quasiparticle poisoning can be the presence of quasiparticle
states in the gap of Nb. As a strong gettering material, Nb may
react with the chemical residues on the substrate and/or gas
impurities inside the vacuum chamber during the deposition process,
and this can result in creation of subgap states. Also, the quality
of the Al oxide tunnel barrier may degrade during the deposition of
the top Nb layer, leading to the formation of a complex interface
containing lower Nb oxides
between the Al/AlO$_x$ and Nb layers and eventually to the
quasiparticle leakage in the junction. Besides Nb island
nonidealities, presence of Nb in the leads may be an extra source of
quasiparticles contributing to the poisoning in the hybrid
structures.
\newline \indent In conclusion, we have presented a
comparative study of the Cooper pair transistors made with either Al
or Nb islands, embedded in tunable electromagnetic environments. The
device with an Al island demonstrates gate charge modulation with
clear $2e$-periodicity in a wide range of environmental impedances
as long as the bath temperature is kept below 340 mK. This suggests
that the Al samples are of good quality and the filtering of the
measurement set-up is sufficient to avoid quasiparticle poisoning.
Contrary to the Al sample, the three similar samples with a Nb
island measured in the same set-up exhibit only $e$-periodicity
under all experimental conditions. Based on our observations, we
attribute the absence of $2e$-periodicity in the CPTs with Nb
islands to the material properties of Nb. Thus the suitability of Nb
as the material for single Cooper pair devices still remains an
issue.

We thank T. Holmqvist for assistance in the measurements and A.
Abdumalikov, Y. Nakamura and M. Watanabe for useful comments. This
work was partly supported by "RSFQubit" FP6 Project of the European
Union and by Japan Science and Technology Agency.

\end{document}